\documentclass[12pt]{article}
\usepackage[utf8]{inputenc}
\usepackage[letterpaper, margin=1in]{geometry}
\usepackage[superscript]{cite}
\usepackage[doublespacing]{setspace}
\usepackage{lineno}
\usepackage{graphicx}
\usepackage{float}
\usepackage{caption}
\usepackage{bm}
\usepackage[tbtags]{amsmath}
\usepackage{booktabs}

\begin{document}


\begin{singlespace}

\noindent \textbf{Abdominal synthetic CT reconstruction with intensity projection prior for MRI-only adaptive radiotherapy}
\newline

\noindent \textbf{Running title:} Abdominal sCT reconstruction
\newline

\noindent Sven Olberg$^{1,2}$, Jaehee Chun$^{4}$, Byong Su Choi$^{1,4}$, Inkyung Park$^{1,4}$, Hyun Kim$^{3}$, Taeho Kim$^{3}$, Jin Sung Kim$^{4*}$, Olga Green$^{3}$, Justin C. Park$^{1}$
\newline

\noindent $^{1}$Medical Artificial Intelligence and Automation (MAIA) Laboratory, Department of Radiation Oncology, University of Texas Southwestern Medical Center, Dallas, TX 75390, USA
\newline

\noindent $^{2}$Department of Biomedical Engineering, Washington University in St. Louis, St. Louis, MO 63110, USA
\newline

\noindent $^{3}$Department of Radiation Oncology, Washington University in St. Louis, St. Louis, MO 63110, USA
\newline

\noindent $^{4}$Department of Radiation Oncology, Yonsei Cancer Center, Yonsei University College of Medicine, Seoul, South Korea
\vfill

\noindent \textbf{*Corresponding Author:}
\newline

\noindent Jin Sung Kim, Ph.D.
\newline

\noindent Department of Radiation Oncology, Yonsei Cancer Center, Yonsei University College of Medicine, Seoul, South Korea \\ 50, Yonsei-ro, Seodaemun-gu \\ Seoul, South Korea \\ T.+82-2-2228-8118, F.+82-2-2227-7823
\newline

\noindent Email: jinsung@yuhs.ac
\newline

\noindent \textbf{Conflicts of interest:} The authors have no conflicts to disclose.
\newline

\noindent \textbf{Data availability:} The patient data used here is not available at this time.
\newline

\noindent \textbf{Acknowledgment:} J.S.K. acknowledges this work was supported by Radiation Technology R\&D program through the National Research Foundation of Korea funded by the Ministry of Science and ICT.
\newpage

\end{singlespace}

\begin{abstract}

\noindent \textbf{Purpose:} Owing to the superior soft tissue contrast of MRI, MRI-guided adaptive radiotherapy (ART) is well-suited to managing interfractional changes in anatomy. An MRI-only workflow is desirable, but producing synthetic CT (sCT) data through paired data-driven deep learning (DL) for abdominal dose calculations remains a challenge due to the highly variable presence of intestinal gas. We present the preliminary dosimetric evaluation of our novel approach to sCT reconstruction that is well suited to handling intestinal gas in abdominal MRI-only ART.

\noindent \textbf{Methods:} We utilize a paired data DL approach enabled by the intensity projection prior, in which well-matching training pairs are created by propagating air from MRI to corresponding CT scans. Evaluations focus on two classes: patients with (1) little involvement of intestinal gas, and (2) notable differences in intestinal gas presence between corresponding scans. Comparisons between sCT-based plans and CT-based clinical plans for both classes are made at the first treatment fraction to highlight the dosimetric impact of the variable presence of intestinal gas.

\noindent \textbf{Results:} Class 1 patients ($n=13$) demonstrate differences in prescribed dose coverage of the PTV of $1.3 \pm 2.1\%$ between clinical plans and sCT-based plans. Mean DVH differences in all structures for Class 1 patients are found to be statistically insignificant. In Class 2 ($n=20$), target coverage is $13.3 \pm 11.0\%$ higher in the clinical plans and mean DVH differences are found to be statistically significant.

\noindent \textbf{Conclusions:} Significant deviations in calculated doses arising from the variable presence of intestinal gas in corresponding CT and MRI scans result in uncertainty in high-dose regions that may limit the effectiveness of adaptive dose escalation efforts. We have proposed a paired data-driven DL approach to sCT reconstruction for accurate dose calculations in abdominal ART enabled by the creation of a clinically unavailable training data set with well-matching representations of intestinal gas.
\newpage 

\end{abstract}

\section{Introduction}

The proliferation of magnetic resonance imaging (MRI)-based radiotherapy delivery systems in recent years has pushed applications of MRI-guided radiation therapy (MRgRT) to the forefront of RT.\cite{Mutic2014, Fallone2014, Keall2014, Fischer-Valuck2017, Raaymakers2017, Wen2018, Pollard2017} The superior soft tissue contrast of MRI compared to that of x-ray computed tomography (CT) improves target delineation in many sites, enabling near real-time motion tracking and management during treatment.\cite{Noel2015, Raaymakers2017} The advantages of MRI also lend themselves well to managing interfractional changes in a patient's anatomy, as MRI setup scans acquired in an adaptive radiotherapy (ART) workflow capture the anatomy of the day without exposing the patient to additional ionizing radiation.\cite{Fischer-Valuck2017} The need for electron density information in dose calculations, however, necessitates that these setup scans be registered to a CT simulation scan that may have been acquired weeks prior to a given treatment fraction.\cite{Schmidt2015} Issues may arise in some cases when the anatomy represented in each scan is incompatible due to changes in the geometry and position of organs of interest, which is a challenge that is especially relevant to gastrointestinal (GI) structures and intestinal gas pockets considered during ART in the abdomen. In these cases, manual contouring is required so that density overrides may be performed in order to approximate in the CT simulation scan the anatomy of the day captured in the MRI setup scan. This conventional clinical approach of performing air-to-water overrides in the CT scan and transferring air contours from the MRI scan relies on well-matching bony anatomy with only the GI structures demonstrating deviations, which are conditions that are not always easily met.

It is this potential for challenge and uncertainty that comes with multi-modal image registration and intensive contouring that makes an MRI-only workflow---one in which MRI is the sole imaging modality used for planning and guidance---an attractive alternative to the conventional MRgRT workflow. The primary challenge in an MRI-only workflow is generating synthetic CT (sCT) data that yields the electron density information used in dose calculations. Many existing approaches to this task have been summarized as belonging to three general classes: atlas-based, voxel-based, and learning-based methods.\cite{Edmund2017, Johnstone2018} Recent investigations have focused primarily on approaches belonging to the last category, namely deep learning (DL)-based approaches in which convolutional neural networks are used to approximate a mapping between MRI and CT images.

Studies in this space have primarily been limited to regions of relatively static anatomy, including the head \& neck\cite{Han2017} and general pelvis.\cite{Arabi2018, Chen2018a, Maspero2018a} More recently, our group has extended these investigations into the thorax, where registration is challenged by pulmonary and cardiac motion, by generating sCT data for use in MRI-only breast RT.\cite{Olberg2019} When considering applications in the more dynamic region of the abdomen, a primary challenge to employing these deep learning strategies becomes readily apparent: the majority of frameworks require paired data for training. The presence and hard-to-characterize motion of intestinal gas gives rise to observable differences in bowel filling and position on numerous time scales: seconds in the course of a single scan, minutes during treatment delivery, and hours between MRI and CT scans used for treatment planning.\cite{Mostafaei2018, Nakamoto2004, Feng2009, Kumagi2009, Corradini2019} This challenge is even more relevant clinically when considering setup scans in MRI-guided adaptive treatments. In these situations, MRI setup scans acquired the day of treatment are registered to the CT simulation scan that may have been acquired weeks previously. When large discrepancies in bowel filling and position exist, significant manual involvement is required in contouring so that density overrides may be performed. In an online adaptive workflow, this time investment represents a potentially significant delay to treatment delivery while the patient remains on-table.\cite{Henke2018} As an alternative to the paired data approaches, one may consider the application of an unpaired style-transfer approach exemplified by CycleGAN in settings in which abundant, unpaired data exists.\cite{Zhu2018} However, the authors of the original CycleGAN paper acknowledge a gap between the paired and unpaired results that is hard or even impossible to close in some settings, especially those in which there exists some inherent ambiguity.\cite{Zhu2018} This conclusion is mirrored in other studies of the sCT reconstruction task; Peng et al.\cite{Peng2020} conclude that the conditional paired approach was ``preferable if high-quality MR-CT pairs were available'' in the nasopharynx, which is another area challenged by the presence of air, and Fu et al.\cite{Fu2020} demonstrate no benefit to adopting the CycleGAN in the upper abdomen for liver cancer patients.

In light of these challenges, existing works on sCT reconstruction in the abdomen have abandoned the DL-based approaches in favor of various classification or thresholding-based approaches wherein manual steps are taken in the process of generating sCT data to account for the presence of air. Bredfeldt et al\cite{Bredfeldt2017} and Hsu et al\cite{Hsu2019} utilized a fuzzy c-means clustering algorithm to classify tissue classes based on multiple MRI volumes captured using different sequences, taking care to threshold image regions where air is expected to be found before applying the classification algorithm. Alternatively, Ahunbay et al\cite{Ahunbay2019} opted to use deformable image registration between the daily MRI and simulation CT scans to transfer electron density information while using thresholding operations in manually contoured regions to identify the presence of air. Guerreiro et al\cite{Guerreiro2019} explored a hybrid atlas and intensity-based conversion algorithm\cite{Korhonen2014, Koivula2016, Koivula2017} in which contoured regions of air were transferred directly from the simulation CT after Hounsfield units (HU) were assigned.

In the present study, we return to paired data-driven DL with a novel hybrid approach to the abdominal sCT reconstruction task enabled by the creation of a training data set that is clinically unavailable. As was previously discussed, the primary barrier to the adoption of many DL-based algorithms in this setting is the requirement for paired training data. Mismatches in the presence of intestinal gas between corresponding MRI and CT scans render the collection of a training data set of sufficient size an impossible task. Considering this and the challenges faced with an unpaired approach, we first utilize automated thresholding and morphological reconstruction operations to identify and propagate regions of air between corresponding MR and CT images, producing a well-matched training data set and avoiding intensive manual contouring as a post-processing step in the clinical setting. We present here the preliminary evaluation of our paired data DL-based approach to sCT reconstruction in the abdomen enabled by the novel utilization of the intensity projection prior with a focus on showcasing the effects of intestinal gas differences on dose calculations. Dosimetric comparisons are made between two classes of test patients: Class 1, consisting of well-matched patients demonstrating little involvement of intestinal gas, and Class 2, consisting of patients characterized by notable differences in the presence of intestinal gas in corresponding MRI and CT scans. Dosimetric accuracy is established using the patients of Class 1 while comparisons of target coverage between the sCT-based plans and the simulation CT-based clinical plans for patients of Class 2 highlight the complications posed by intestinal gas during MRI-only ART in the abdomen.

\section{Materials and Methods}

\subsection{Patient population}

Data sets used in the present study were retrospectively collected from a population of pancreatic cancer patients previously treated at our institution using MRgRT. In each case, patients underwent CT and MRI simulation scans prior to treatment planning. Scans were acquired in treatment position using an Alpha Cradle (Smithers Medical Products Inc., North Canton, OH) with no additional immobilization devices. The nominal prescription in the selected population was a total dose of 50 Gy delivered to 95\% of the PTV in 5 fractions. A small number of cases deviated from this prescription with total prescribed doses ranging from 30--40 Gy delivered in 5 fractions. For the nominal prescription, satisfying dose-volume constraints of $<$0.5 cm$^3$ at 36 Gy for notable structures including the duodenum, stomach, small bowel, and large bowel was prioritized over target coverage. In any case in which one of these constraints was violated, the calculated dose was normalized to satisfy the violated constraint. The same normalization applied in the clinical case was also applied in the evaluation of our proposed method as discussed later.

\subsection{Training and testing data}

A total of 89 patient data sets (one pair of corresponding MRI and CT scans per patient) were used in the present study, randomly assigned in the following splits: 53 train / 3 valid / 33 test. Validation of the framework in the sCT reconstruction task has been previously carried out by our group.\cite{Olberg2019} The 33 test patients were qualitatively subdivided prior to testing into the two classes mentioned previously: 13 well-matched patients in Class 1 and 20 patients characterized by notable discrepancies in bowel filling in Class 2. CT simulation scans acquired using a dedicated simulation machine (Brilliance CT, Philips Medical Systems, Andover, MA) were registered with 0.35 T MRI scans (nominally 276 $\times$ 276 $\times$ 80 matrix, 1.63 $\times$ 1.63 $\times$ 3 mm$^3$) acquired using the MRIdian system (ViewRay Inc., Oakwood Village, OH) with a bSSFP sequence before being exported for pre-processing. Processing yielded a training data set of 2017 paired images, which were padded to dimensions of 520 $\times$ 520 before training via 320 $\times$ 320 random crops. The framework was trained for 1500 epochs with a batch size of 1 using TensorFlow\cite{tf2015} v1.7.0 in Python running on a 12 GB Titan Xp GPU (NVIDIA, Santa Clara, CA).

\subsection{Pre-processing}

A primary challenge in training a generative model to solve an image-to-image translation task such as this is constructing a set of training data consisting of well-matched pairs of images. This process becomes even more complicated in the abdomen when the variable presence of intestinal gas must be considered. Corresponding MRI and CT scans used in treatment planning may demonstrate notable mismatches in bowel filling and position that present a barrier to accurate dose calculations. While these mismatches are traditionally handled during the treatment planning process in the clinical setting through intensive manual contouring to enable electron density overrides, we opt to eliminate this burden from the clinical setting through the creation of an intensity projection prior. Here we adopt a novel approach to data augmentation for paired data DL applications in which incompatible representations of intestinal gas in corresponding MRI and CT scans are made compatible through the propagation of air from MRI to CT images. The handling of intestinal gas proceeded in the following steps:

(1) Corresponding MRI and CT scans were rigidly registered in the ViewRay treatment planning system (ViewRay Inc., Oakwood Village, OH) to achieve a gross alignment, primarily of bony anatomy.

(2) Using automated thresholding and morphological reconstruction operations, regions of air in each scan were identified. For CT images, Otsu's method \cite{Otsu1979} was used to compute a single threshold value with which the image could be quantized to produce a body mask that excludes most notably the couch. Within this mask, regions of air were thresholded with a histogram shape-based method by selecting intensities falling within an offset (defined as 7 bin widths here) around the lower mode. Finally, binary erosion and dilation operations were performed to eliminate small, noisy regions and produce a smoother segmentation, respectively. For MR images, a similar approach was taken to produce a body mask. In identifying regions of air within this mask, extra precautions were required to avoid selecting low-signal regions not containing air (e.g. vertebral body). To this end, the erosion of a quantized image based on a five-level thresholding achieved with Otsu's method was used as the basis for the morphological reconstruction \cite{Vincent1993} of segmented regions within the abdominal cavity as defined by the space surrounded by the body wall and excluding the region around the vertebral body.

(3) These regions in the CT images were infilled with realistic texture via harmonic inpainting to produce a CT image with no air-containing regions. \cite{Schonlieb2015, Parisotto2016}

(4) Regions of air identified in the MR images in step (2) are then propagated to the corresponding CT images. As such, regions of air originally represented in the CT image but not the corresponding MR image maintain the infilled texture from step (3). Finally, the scans are deformably registered by way of a statistical dependence measure algorithm \cite{Shi2013}.



\subsection{Model and loss formulation}

The task of sCT reconstruction viewed as a forward mapping from MRI to CT has been previously discussed\cite{Olberg2019}. Briefly, the goal in establishing a generative model is to estimate a suitable operator that maps from MRI to CT, which is challenged by the many-to-one correspondence of pixel intensities between the two modalities. We approach the image-to-image translation task using a generative adversarial network (GAN) framework consisting of two competing networks: (1) a generative model $G$ that produces sCT samples residing in the same space as true CT data and (2) a discriminator $D$ that attempts to distinguish between samples generated by $G$ and true samples. During training, $G$ and $D$ undergo alternating minimization steps of their respective loss function, each of which depends on the generalized definition of sigmoid cross entropy loss\cite{tf2015}:

\begin{equation}
    L = \vec{x} - \vec{x} * \vec{z} + \log{(1 + \exp{(-\vec{x})})},
    \label{eq:scel}
\end{equation}

where $\vec{x}$ is the true or predicted image logits computed by $D$ and $\vec{z}$ is a label corresponding to true ($\vec{1}$) or predicted ($\vec{0}$). Using this definition of sigmoid cross entropy loss, the generator loss function $g_{loss}$ is defined by

\begin{equation}
g_{loss} = L_{adv} + l_{mae},
\end{equation}

where the adversarial loss $L_{adv}$ is the sigmoid cross entropy loss (Eq. \ref{eq:scel}) with predicted image logits $\vec{x}$ assigned a true label ($\vec{z} = \vec{1}$) and the mean absolute error (MAE) loss $l_{mae}$ is the mean of the absolute difference between true images $I_{true}$ and predicted images $I_{pred}$:

\begin{equation}
l_{mae} = \text{mean} \left( \left| I_{pred} - I_{true} \right| \right).
\end{equation}

Minimizing such a loss formulation yields synthetic images that are computed as true images by the discriminator through the adversarial term $L_{adv}$ while also maintaining pixel-wise agreement between the generated and true images through the MAE term $l_{mae}$.

While true labels are assigned to predicted images in $g_{loss}$, the discriminator aims to correctly identify true and predicted images. As such, the discriminator loss function $d_{loss}$ depends only on the sigmoid cross entropy loss:

\begin{equation}
d_{loss} = L_{pred} + L_{true},
\end{equation}

where $L_{true}$ and $L_{pred}$ are the sigmoid cross entropy loss (Eq. \ref{eq:scel}) with true or predicted image logits $\vec{x}$ and corresponding labels $\vec{z} = \vec{1}$ or $\vec{z} = \vec{0}$, respectively. While $G$ strives to generate outputs computed as true images by $D$ through the adversarial loss term $L_{adv}$ with predicted image logits assigned a true label, the expected true and false labels are instead used for the true and predicted CT images in the formulation of $d_{loss}$.

Trainable paramaters describing the various operations in each of the competing networks are optimized through alternating minimization steps of the respective loss functions. For $d_{loss}$, TensorFlow's\cite{tf2015} gradient descent optimizer is used with an initial learning rate of 0.00002. In the following minimization step, $g_{loss}$ is minimized utilizing the Adam gradient-based stochastic optimization algorithm\cite{Kingma2017} with an initial learning rate of 0.0002, $\beta_1 = 0.7$, $\beta_2 = 0.999$, and $\hat{\epsilon} = 10^{-8}$. In both cases, learning rates decay every 10000 steps subject to a staircase exponentional function with a decay rate of 0.95.

\subsection{Network architecture}

\subsubsection{Generator}

\begin{figure}[h!]
	\centering
	\includegraphics[width=\textwidth]{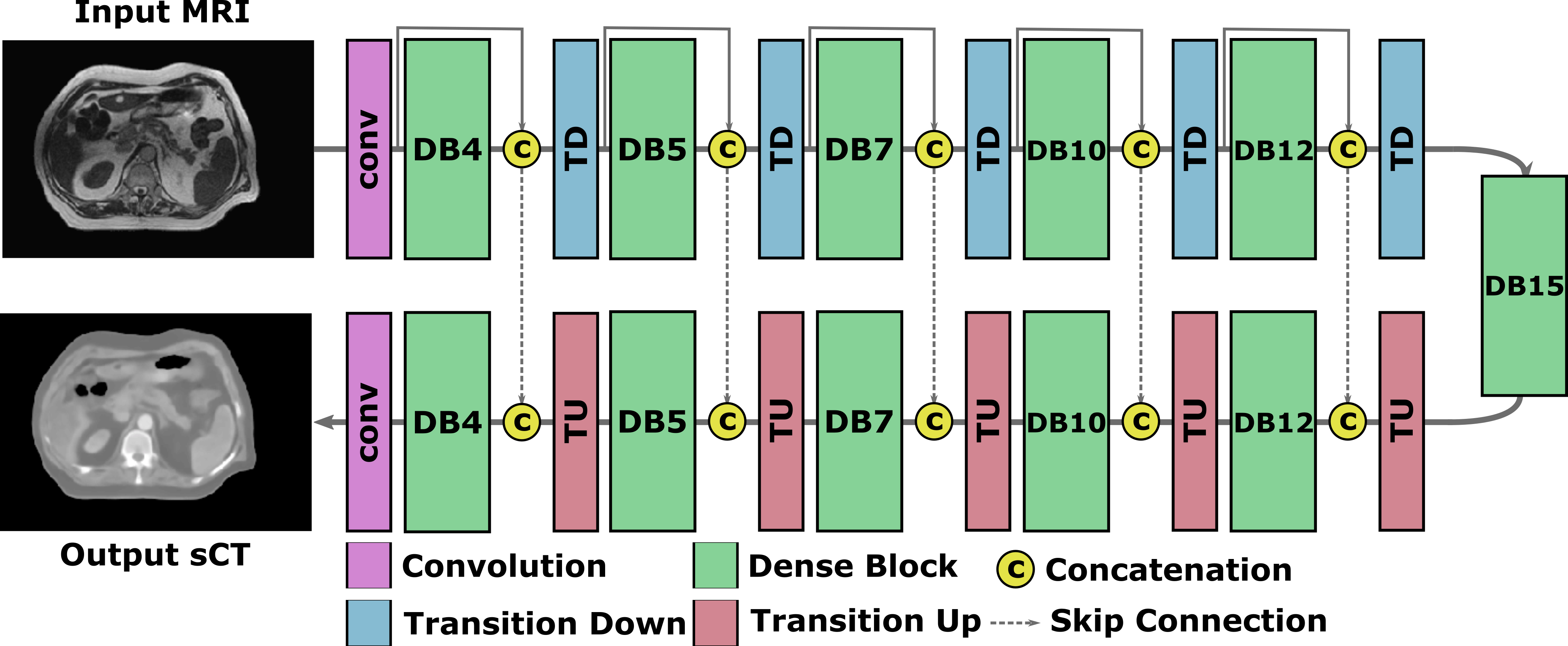}
	\caption{The DenseNet architecture. Individual dense blocks are arranged to form a stacked encoder-decoder U-net structure. DB$n$ denotes a dense block consisting of $n$ intermediate layers. The input MR image is encoded as a set of feature maps that grows progressively deeper as it travels through the encoder layers. Transpose convolution operations in the decoder recover the input spatial resolution, reconstructing details in the output sCT image with the aid of skip connections that transfer structural information from the encoder.}
	\label{fig:fcdn}
\end{figure}

The fully convolutional DenseNet\cite{Jegou2017} employed here, illustrated in Figure \ref{fig:fcdn}, consists of individual dense blocks arranged to form a stacked encoder-decoder U-net\cite{Ronneberger2015} structure. The constituent dense blocks, which resemble residual blocks\cite{He2015} in that intermediate feature maps are iteratively concatenated, are built using the following components: Batch Normalization (BN), ReLU activation, $3 \times 3$ same convolution, and dropout with probability $p = 0.2$. The growth rate $k$ of the layer ($k = 16$ in this case) dictates the number of feature maps computed by each layer. Feature maps computed by each of these intermediate layers are iteratively concatenated to form the output of the dense block itself, granting a degree of convergence-aiding supervision due to the short paths to all feature maps in an architecture that is ultimately quite parameter efficient\cite{Jegou2017}. Transition down (TD) operations in the encoder path, which serve to reduce the dimensionality of feature maps, consist of BN, followed by ReLU activation, $1 \times 1$ convolution, dropout with probability $p = 0.2$, and $2 \times 2$ max pooling with stride $2$. To recover the input dimensions on the decoder path, transition up (TU) layers perform $3 \times 3$ transpose convolution with stride $2$. Skip connections between corresponding layers of the encoder and decoder sides of the network transfer structural information that aids in the reconstruction of fine detail as the full input resolution is recovered along the decoder path.

\subsubsection{Discriminator}

The architecture of the discriminator is unchanged from the previous application to sCT generation in the breast\cite{Olberg2019}. $D$ is a straightforward encoder consisting of five convolutional layers that ultimately applies the sigmoid function to yield the probability of the evaluated image being a true CT image.

\subsection{Evaluation}

The proposed approach to sCT reconstruction is evaluated in two primary ways, each with a focus on the two classes of patients previously discussed. Pixel-wise image comparisons are made between true CT images and reconstructed sCT images for patients belonging to each class using the MAE measured in regions within the body contour not containing air:

\begin{equation}
    \text{MAE} = \frac{\sum_{i=1}^{n}\left| CT_i - sCT_i \right|}{n},
\end{equation}

where $n$ is the number of pixels not containing air in both the CT reference image and the generated sCT image. These image comparisons are made for sCT outputs of both the model proposed here and a ``blind'' model trained with the same patient data only without the pre-processing treatment of regions of air, which represents the conventional DL-based approach to the present problem. Considering the fact that the reference CT image for patients of Class 2 may be largely incompatible with the corresponding MRI image, we also evaluate the degree of overlap of regions of air in input MR images and reconstructed sCT outputs of each model using the Dice similarity coefficient (DSC):

\begin{equation}
\text{DSC} = \frac{2 \left| X \cap Y \right|}{\left| X \right| + \left| Y \right|},
\end{equation}

where $X$ and $Y$ are the sets of pixels in air masks of an MR image and corresponding sCT image.

A subsequent dosimetric evaluation compares dose distributions calculated at the first treatment fraction in simulation CT-based clinical plans to those recalculated using sCT-derived electron density information and the same optimization parameters used in the clinical plans for each of the 33 test patients. Full dose-volume histograms (DVHs) for the target and surrounding tissues of well-matched patients in Class 1 are used to first establish the baseline accuracy of the proposed approach to sCT reconstruction. The same comparison is made for patients in Class 2 demonstrating notable discrepancies in the presence of intestinal gas between corresponding MRI and CT scans to explore the effect of these discrepancies. For both patient classes, we examine differences in prescribed dose coverage of the PTV between the clinical CT-based plans and the proposed sCT-based plans. The 3D gamma index with a 3\%/3 mm criterion is computed to evaluate agreement between the CT-based and sCT-based dose distributions for patients in each class.\cite{Low1998} Additionally, mean DVH differences for each structure of interest are computed and evaluated for statistical significance in each patient class. For patients of Class 2, the full width at half maximum (FWHM) of the profile of the difference in calculated target coverage is used to evaluate the uncertainty in high-dose coverage of the target due to the involvement of intestinal gas. Dose calculations in each case were performed using the ViewRay treatment planning system integrated Monte Carlo algorithm in the presence of a magnetic field with a dose grid resolution of 0.3 cm and calculation uncertainty of 1\%.

\section{Results}

\subsection{Image comparison}

Completing 1500 epochs during training required 126 hours in total. At deployment, inference requires approximately 0.26 s/slice.

\begin{figure}[h!]
	\centering
	\includegraphics[width=\textwidth]{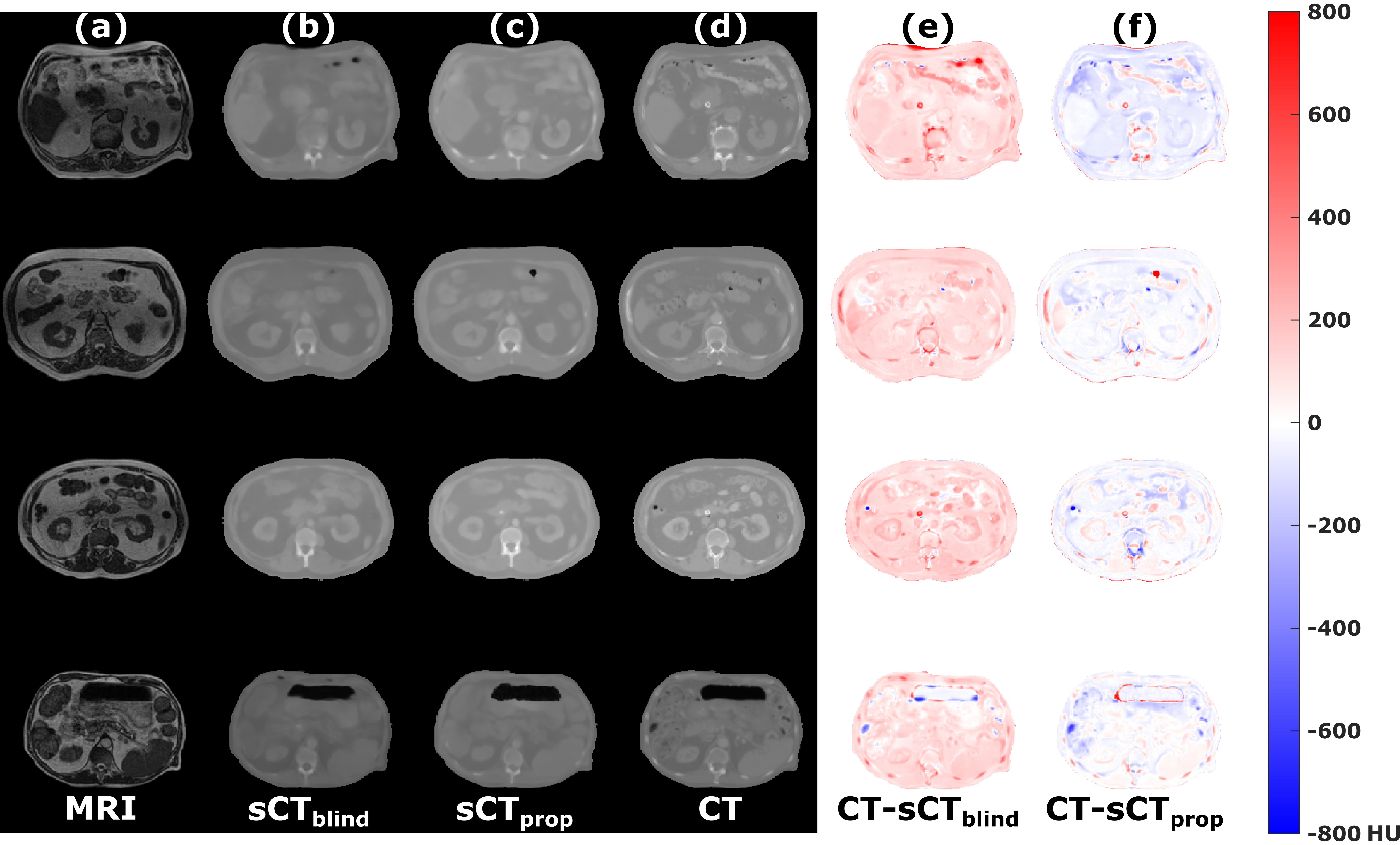}
	\caption{Image comparisons for representative slices of well-matched patients of Class 1. Input MR images (a), output sCT images for the blind model (b) and proposed model (c), true CT images (d), and difference maps (e-f) between the true CT images and generated sCT images for the blind and proposed model, respectively. Values in the difference maps are in units of HU. The final row illustrates the rare case of a relatively well-matched slice with a notable presence of intestinal gas.}
	\label{fig:diff-no-gas}
\end{figure}

Each of the following comparisons shows an input MR image along with the sCT reconstruction produced by the blind and proposed models and the corresponding true CT image. Difference maps illustrate differences in pixel intensities between the true CT image and sCT image in units of HU. Axial slices for representative patients belonging to Class 1 are displayed in Figure \ref{fig:diff-no-gas}. Difference maps (Figure \ref{fig:diff-no-gas}e-f) between the true CT image and the sCT reconstruction for each patient show a general agreement in the bulk of the soft tissue represented in each image in the case of the proposed model and a failure to accurately reproduce HU values in the case of the blind model. This is reflected in the MAE values computed for patients of Class 1; the blind and proposed approaches achieve values of $143 \pm 29$ HU and $90 \pm 29$ HU, respectively. Also included in the last row of this comparison is an example of the relatively rare case of a notable presence of intestinal gas that is well-matched in corresponding MRI and CT scans. 

\begin{figure}[h!]
	\centering
	\includegraphics[width=\textwidth]{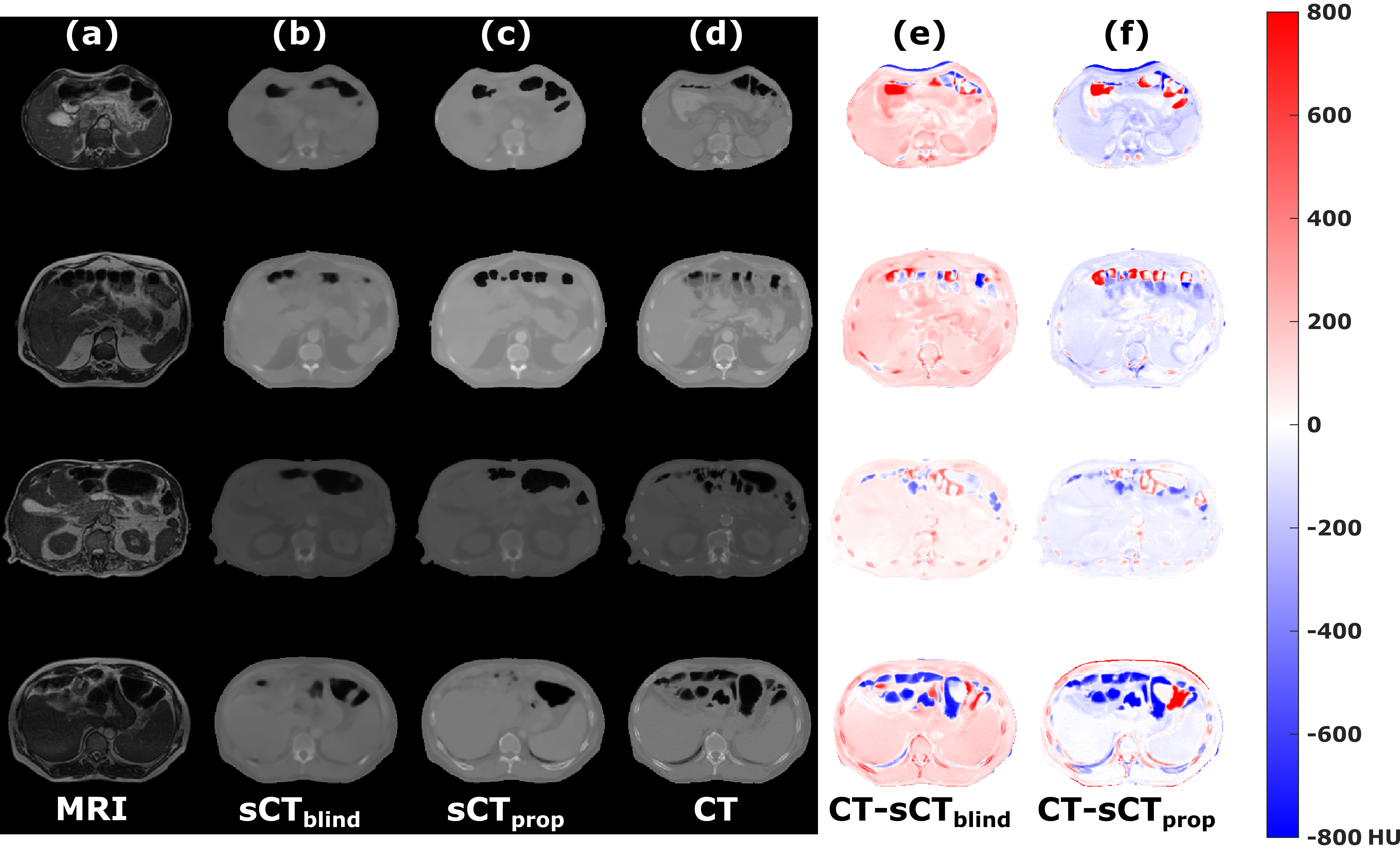}
	\caption{Image comparisons for representative slices of patients in Class 2 characterized by notable differences in the presence of intestinal gas between corresponding MRI and CT scans. Input MR images (a), output sCT images for the blind model (b) and proposed model (c), true CT images (d), and difference maps (e-f) between the true CT images and generated sCT images for the blind and proposed model, respectively. Values in the difference maps are in units of HU.}
	\label{fig:diff-gas}
\end{figure}

In contrast, Figure \ref{fig:diff-gas} shows the same comparison made for patients of Class 2 in which notable discrepancies in the presence of intestinal gas between corresponding MRI and CT scans are observed. These discrepancies give rise to pixel-wise disagreements of the order of $\pm$ 800 HU in the involved gas-containing regions. At the same time, the failure on the part of the blind model to produce accurate HU values in regions of soft tissue is observed. The increased likelihood of discrepancies in soft tissue positions between corresponding scans in patients of Class 2 causes an increase in the MAE computed in regions not containing air in either image: up to $164 \pm 41$ HU for the blind model and $112 \pm 41$ HU for the proposed model.

The overlap of regions of air in MR images and the corresponding sCT reconstructions was evaluated in a total of 158 images from the four patients included in Figure \ref{fig:diff-gas} using the DSC. The average DSC improved from 0.56 for the blind model to 0.80 for the proposed model.

\subsection{Dosimetric evaluation}

\begin{figure}[h!]
	\centering
	\includegraphics[width=\textwidth]{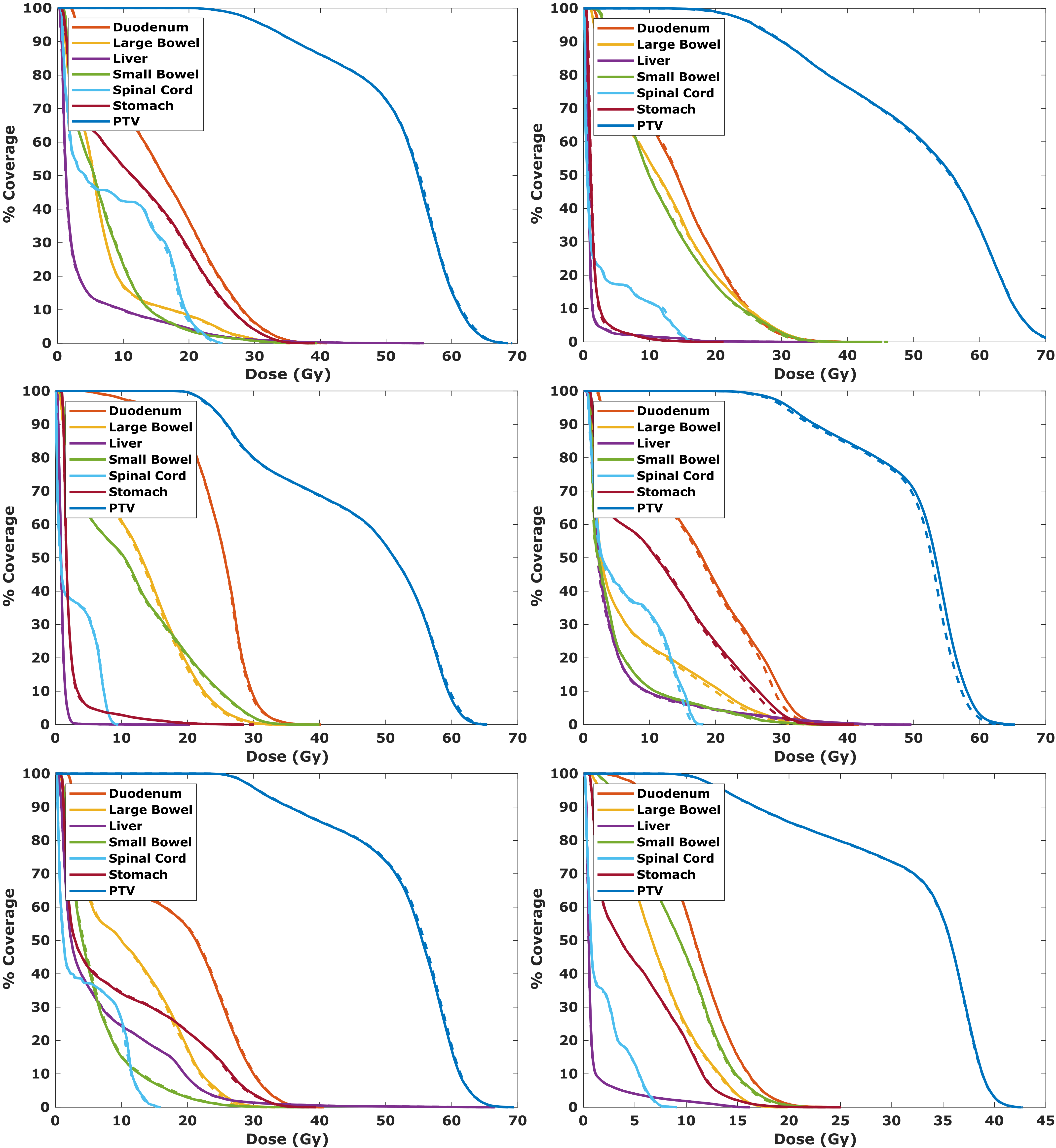}
	\caption{Representative DVHs for well-matched test patients of Class 1 comparing the CT-based clinical plans (dashed) and sCT-based plans (solid) recalculated using the same plan parameters. The prescribed dose was 50 Gy in all but the last case.}
	\label{fig:dvh-class1}
\end{figure}

Similarly, the dosimetric evaluation of the proposed approach to sCT reconstruction focused on the two distinct classes of patients. In both cases, optimization parameters selected in the CT-based clinical plans were used to recalculate dose distributions based on electron density information derived from the generated sCT images. Used as a baseline point of reference to establish the dosimetric accuracy of the proposed reconstructions, the thirteen well-matched patients of Class 1 demonstrate differences in the prescribed dose coverage of the PTV ($V_{100}$) of $1.3 \pm 2.1\%$ between CT-based clinical plans and the sCT-based plans with a gamma pass rate of $98.3 \pm 1.3\%$ using 3\%/3 mm criterion. The representative DVHs for patients belonging to Class 1 shown in Figure \ref{fig:dvh-class1} demonstrate close agreement in calculated target coverage and doses to surrounding tissues between the CT-based clinical plans and the sCT-based recalculations.

\begin{figure}[h!]
	\centering
	\includegraphics[width=\textwidth]{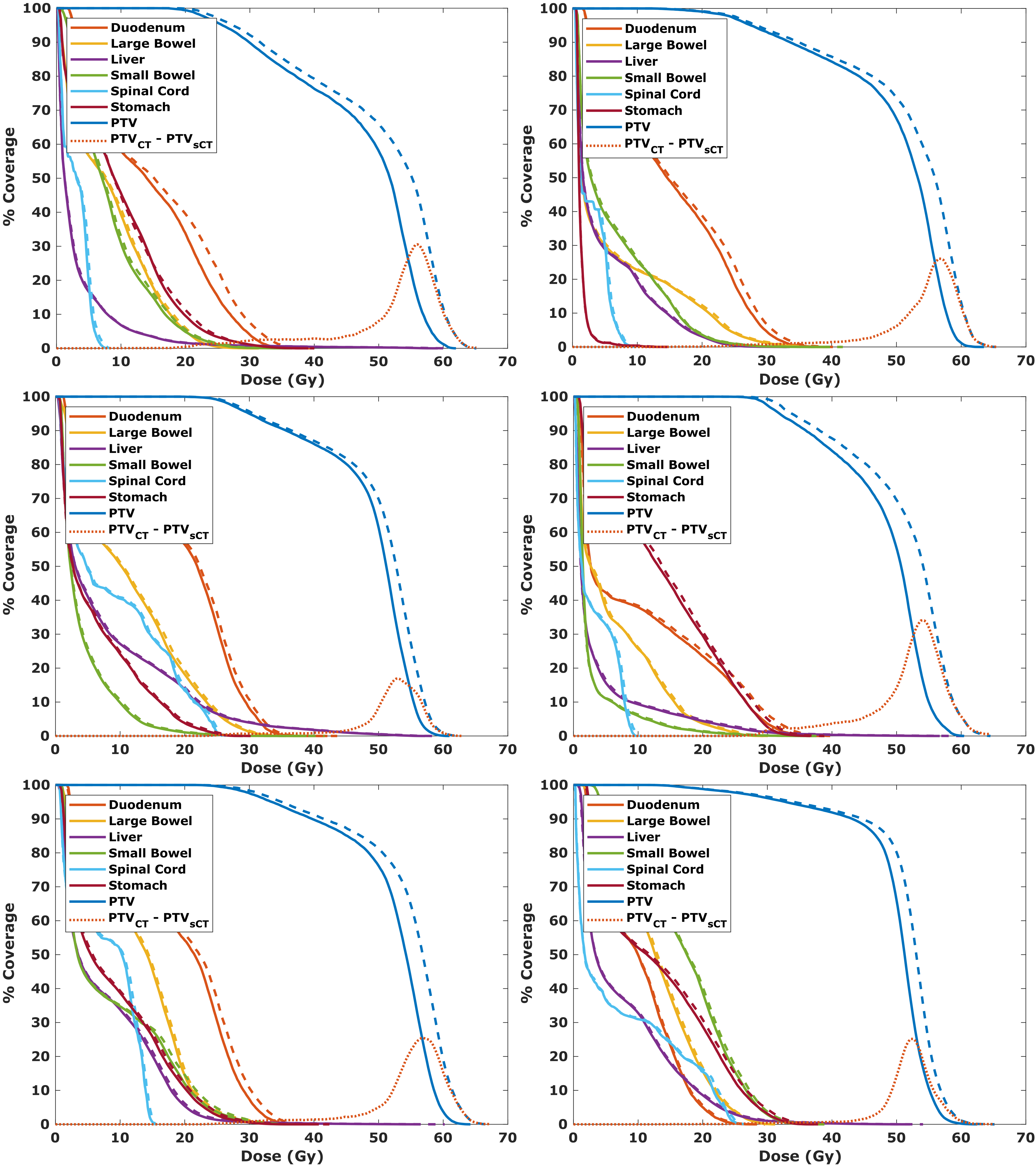}
	\caption{Representative DVHs for patients of Class 2 characterized by notable differences in the presence of intestinal gas between corresponding MRI and CT scans comparing the CT-based clinical plans (dashed) and sCT-based plans (solid) recalculated using the same plan parameters. Differences in calculated target coverage are plotted at each point (dotted) to yield an approximately Gaussian curve. The FWHM of these difference profiles for all patients of Class 2 covers an average range of 51.4--58.2 Gy.}
	\label{fig:dvh-class2}
\end{figure}

\begin{figure}[h!]
	\centering
	\includegraphics[width=\textwidth]{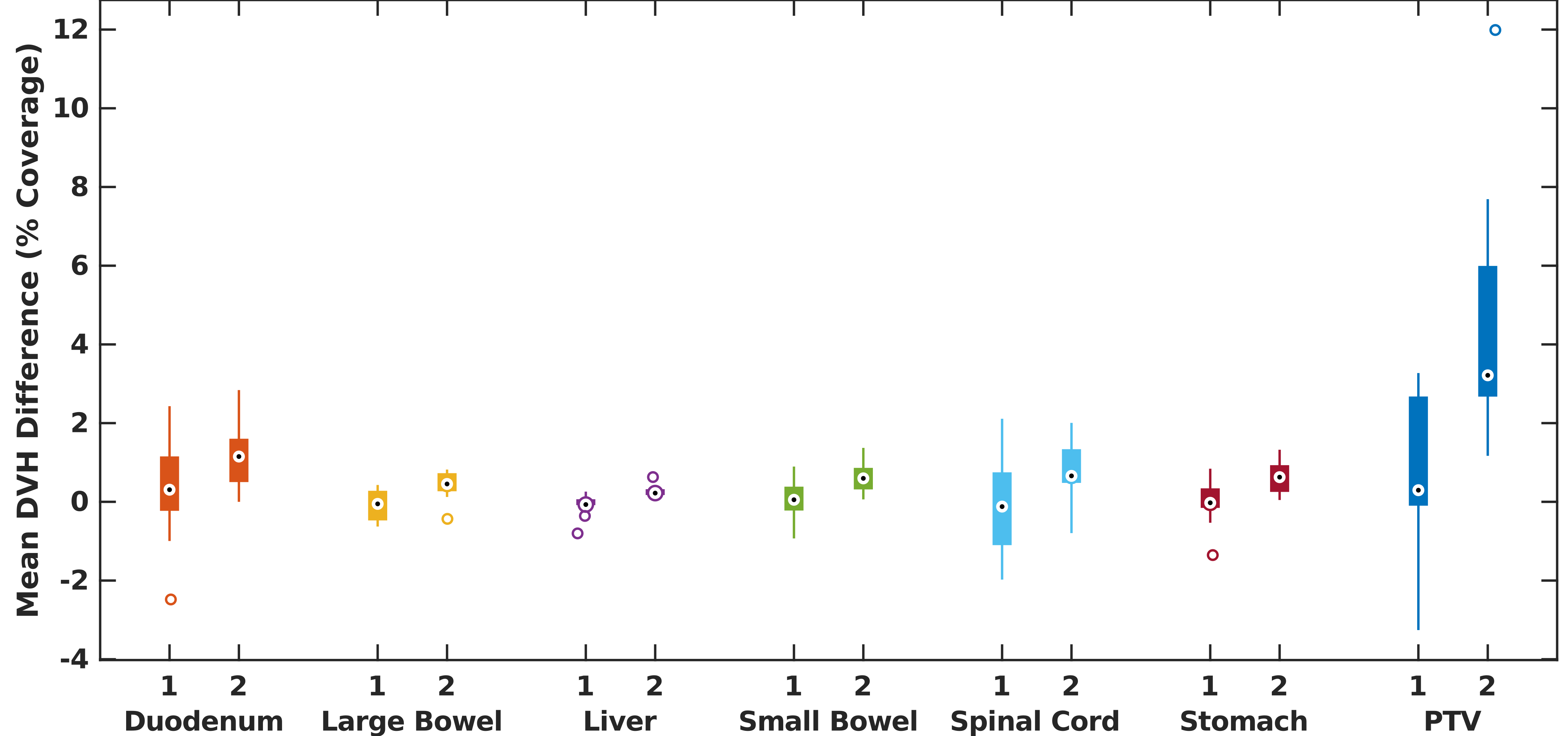}
	\caption{Summary of mean DVH differences between CT-based clinical plans and sCT-based recalculated plans for each structure of interest in well-matched patients of Class 1 ($n=13$) and patients of Class 2 ($n=20$) characterized by notable differences in the presence of intestinal gas between corresponding MRI and CT scans.}
	\label{fig:dvh_diff}
\end{figure}

For the twenty poorly-matched patients of Class 2, notable discrepancies in the representation of intestinal gas between corresponding MRI and CT scans result in sizeable and variable differences in PTV $V_{100}$ coverage: $13.3 \pm 11.0\%$ on average. Due to these differences, the gamma pass rate is reduced to $93.9 \pm 9.8\%$ using the same 3\%/3 mm criterion. These differences in target coverage, along with small discrepancies in the dose to closely involved tissues like the duodenum, are observable in the representative DVHs included in Figure \ref{fig:dvh-class2}. Also plotted in Figure \ref{fig:dvh-class2} is the difference in target coverage at each point, which yields an approximately Gaussian profile. The FWHM of this profile for all patients in Class 2 covers an average range of 51.4(SD=1.3)--58.2(1.6) Gy. Mean DVH differences between the CT-based clinical plans and the sCT-based plans for all structures of interest are plotted in Figure \ref{fig:dvh_diff} for patients of each class. For patients in Class 1, differences between the CT-based and sCT-based plans are computed to be statistically insignificant (distributed with a median of zero) using a two-sided Wilcoxon signed rank test\cite{Wilcoxon1945} for each of the duodenum (p=.34), large bowel (p=.62), liver (p=.52), small bowel (p=.38), spinal cord (p=.91), stomach (p=.91), and PTV (p=.20). For patients in Class 2, differences were instead shown to be statistically significant (not distributed with a median of zero) using a two-sided Wilcoxon signed rank test for each of the duodenum (p=.002), large bowel (p=.007), liver (p$<$.001), small bowel (p$<$.001), spinal cord (p=.01), stomach (p$<$.001), and PTV (p$<$.001).

\section{Discussion}

The potential value of generating synthetic CT data for MRI-only ART in the abdomen is multifaceted. Although therapeutic gains may be achieved, adopting an online adaptive workflow introduces additional time burdens to the process of treatment delivery including re-contouring, re-planning, and quality assurance---all of which must occur while the patient remains on-table. Re-contouring, which must be undertaken to accommodate changes in both normal tissue volumes and also the variable presence of intestinal gas that is our current focus, represents a significant portion of the total on-table time per fraction: up to 24 minutes in the worst case.\cite{Henke2018} Furthermore, this burdensome approach of performing air-to-water overrides in the CT scan and transferring contoured regions of air from the MRI scan relies on well-matching bony anatomy, which is a condition that is not always easily met. By utilizing the proposed approach to sCT reconstruction explored here in which the focus was placed on producing a clinically unavailable data set of well-matched representations of intestinal gas, one is able to rapidly produce (0.26 s/slice) sCT data in the clinical setting that accurately reflects both the presence of intestinal gas shown in a patient's daily MRI scan and HU values present in a true CT scan. In this way, one of the primary concerns necessitating re-contouring in the adaptive setting is potentially eliminated. When paired with an auto-contouring strategy designed for MRI-guided ART\cite{Fu2018}, the time burden associated with re-contouring may become negligible.

Another primary motivation in adopting an adaptive workflow is to achieve dose escalation under shifting anatomic conditions.\cite{Henke2018b} Plan adaption is often performed to increase OAR sparing while also increasing target coverage.\cite{Henke2018} The observed underdosing of the target for patients of Class 2 characterized by mismatched representations of intestinal gas is especially relevant in these scenarios when dose escalation is a specific aim of pursuing plan adaption. The higher calculated target coverage in clinical plans in which mismatches between the planning CT and setup MRI are not accounted for compared to the sCT-based plans and uncertainty in high-dose regions (Figure \ref{fig:dvh-class2}) represent a barrier to any escalation that is pursued in these scenarios. In the present study, we have explored this effect at the first treatment fraction for 33 test patients, demonstrating that even at the first fraction, a non-negligible portion of the patient population may experience uncertainties in simulation CT-based dose calculations due to the involvement of intestinal gas. This concern becomes even more important when considering the accumulation of dose over the course of a treatment in which each fraction is adapted and these differences accrue. However, it is important to note that the DVHs computed for a given treatment fraction and presented in the ViewRay treatment planning system are full rather than fractional DVHs. Considering this, the DVHs examined in this study convey the overall effect in the case that the magnitude of the discrepancies observed at the first fraction are carried forward through each subsequent fraction. Considering the hard-to-characterize nature of gas motion, we do not examine here whether there exists some sort of interplay effect throughout the course of treatment.

We acknowledge a number of additional limitations to the present study. First and foremost, the image-to-image translation approach employed here struggles fundamentally in a situation in which the anatomy represented in corresponding training image pairs differs. Although we took care in the present study to propagate regions of air, geometric differences in soft tissues surrounding the target are not always handled sufficiently by multi-modal deformable image registration. The model is reasonably robust to the variations present in the training data set, but struggles to faithfully reconstruct the most dynamic tissues and structures. Style-transfer methods that do not rely on matched pairs of training data as exemplified by CycleGAN\cite{Zhu2018} may be of particular use in this application to overcome the limitations of multi-modal deformable image registration. The implementation of an unpaired approach may still benefit from semi-supervised data produced in the manner described here to overcome the limitations of the unpaired approach in cases with inherent ambiguity, as is the case for MRI signal intensity. A second limitation of the trained model is reconstruction inaccuracies at the superior and inferior extremes of a patient's image stack that stem from the make-up of the training data set. While every patient data set was roughly centered on the target and surrounding tissues, slices containing views of the lungs and diaphragm or inferior portions of the abdomen were not as equally represented. These reconstruction inaccuracies may be ameliorated by adopting a more robust 3D network architecture over the relatively lightweight 2D architecture at the expense of heavily increased memory usage, which may not be a feasible trade-off in all settings. An additional concern regarding MR image quality in the abdominal sCT reconstruction task is the issue of susceptibility artifacts in the GI tract. The effects of these artifacts are lessened due to the lower field strength of the MRgRT platform utilized here,\cite{Ginn2017}  but certain circumstances involving the ingestion of fortified foods prior to treatment---although not encountered in the present study---have been reported and are thus an important consideration in this application.\cite{Green2018} 

An inherent challenge in this space is the issue of image evaluation when the underlying premise is that the ``ground truth'' simulation CT data is incompatible with the setup MRI data used due to the involvement of intestinal gas and the motion of GI structures. As such, the image comparisons made here are imperfect comparisons and we instead rely more heavily on the improvement in the representation of air in our sCT reconstructions as measured by the DSC considering that is the primary focus of this work. Nonetheless, a comparison to other methods is warranted. Ahunbay et al.\cite{Ahunbay2019} achieved an MAE of approximately 25 HU in the abdomen using a method entirely reliant of multiple deformations of true CT images. Closer to the realm of true image synthesis, multiple atlas-based techniques have reported values ranging from 40--200 HU for sites including the pelvis, cranium, and general torso.\cite{Guerreiro2019, Edmund2017, Johnstone2018} In the MAE values we report, we do not distinguish between regions of soft tissue and bone, but we do compare competitively to existing, slower methods.

Finally, the separation of test patients into two separate classes performed here relied on the qualitative assessment of the relative involvement of intestinal gas and the degree to which representations of intestinal gas matched between corresponding MRI and CT scans. In some cases, the characterization of the patient was clear---there were easily observable discrepancies or gas was entirely uninvolved---but the categorization was more challenging in other cases. As such, the furthering of this and related work would benefit from some quantitative approach to the characterization of the involvement and similarity of representations of intestinal gas that would in turn enable the exploration of trends in distinct patient groups.

\section{Conclusions}

The approach to sCT reconstruction in the abdomen evaluated here highlights the challenges posed by the presence of intestinal gas throughout the MRI-guided ART workflow. Eliminating the burden of handling intestinal gas from the clinical setting through the creation of a clinically unavailable training data set for training a paired data generative model offers the potential to streamline a time-intensive portion of the adaptive treatment workflow. These time savings are gained while also enabling accurate dose calculations in adaptive treatments despite the variable presence of intestinal gas at each stage of treatment planning and delivery during MRI-only ART in the abdomen.

\begin{singlespace}

\bibliographystyle{ama}
\bibliography{sct-abd-refs}

\begin{thebibliography}{10}
\expandafter\ifx\csname urlstyle\endcsname\relax
  \providecommand{\doi}[1]{doi:\discretionary{}{}{}#1}\else
  \providecommand{\doi}{doi:\discretionary{}{}{}\begingroup
  \urlstyle{rm}\Url}\fi

\bibitem{Mutic2014}
Mutic S, Dempsey JF.
\newblock The {ViewRay} system: Magnetic resonance–guided and controlled
  radiotherapy.
\newblock \emph{Semin Radiat Oncol}. 2014;\hspace{0pt}24(3):196--199.
\newblock \doi{10.1016/j.semradonc.2014.02.008}

\bibitem{Fallone2014}
Fallone BG.
\newblock The rotating biplanar linac-magnetic resonance imaging system.
\newblock \emph{Semin Radiat Oncol}. 2014;\hspace{0pt}24:200--202.
\newblock \doi{10.1016/j.semradonc.2014.02.011}

\bibitem{Keall2014}
Keall PJ, Barton M, Crozier S.
\newblock The {Australian} magnetic resonance imaging-linac program.
\newblock \emph{Semin Radiat Oncol}. 2014;\hspace{0pt}24:203--206.
\newblock \doi{10.1016/j.semradonc.2014.02.015}

\bibitem{Fischer-Valuck2017}
Fischer-Valuck BW, Henke L, Green O, et~al.
\newblock Two-and-a-half-year clinical experience with the world's first
  magnetic resonance image guided radiation therapy system.
\newblock \emph{Adv Radiat Oncol}. 2017;\hspace{0pt}2(3):485--493.
\newblock \doi{10.1016/j.adro.2017.05.006}

\bibitem{Raaymakers2017}
Raaymakers BW, J{\"{u}}rgenliemk-Schulz IM, Bol GH, et~al.
\newblock First patients treated with a 1.5 {T} {MRI-Linac}: clinical proof of
  concept of a high-precision, high-field {MRI} guided radiotherapy treatment.
\newblock \emph{Phys Med Biol}. 2017;\hspace{0pt}62(23):L41--L50.
\newblock \doi{10.1088/1361-6560/aa9517}

\bibitem{Wen2018}
Wen N, Kim J, Doemer A, et~al.
\newblock Magnetic resonance guided linear accelerator for stereotactic
  radiosurgery treatment.
\newblock In \emph{Proceedings of the American Society for Radiation Oncology}.
  p. E479.
\newblock \doi{j.ijrobp.2018.07.1369}

\bibitem{Pollard2017}
Pollard JM, Wen Z, Sadagopan R, Wang J, Ibbott GS.
\newblock The future of image-guided radiotherapy will be {MR} guided.
\newblock \emph{Br J Radiol}. 2017;\hspace{0pt}90(1073):20160667.
\newblock \doi{10.1259/bjr.20160667}

\bibitem{Noel2015}
Noel CE, Parikh PJ, Spencer CR, et~al.
\newblock Comparison of onboard low-field magnetic resonance imaging versus
  onboard computed tomography for anatomy visualization in radiotherapy.
\newblock \emph{Acta Oncol (Madr)}. 2015;\hspace{0pt}54(9).
\newblock \doi{10.3109/0284186X.2015.1062541}

\bibitem{Schmidt2015}
Schmidt MA, Payne GS.
\newblock Radiotherapy planning using {MRI}.
\newblock \emph{Phys Med Biol}. 2015;\hspace{0pt}60(22):R323--R361.
\newblock \doi{10.1088/0031-9155/60/22/R323}

\bibitem{Edmund2017}
Edmund JM, Nyholm T.
\newblock A review of substitute {CT} generation for {MRI}-only radiation
  therapy.
\newblock \emph{Radiat Oncol}. 2017;\hspace{0pt}12(1):28.
\newblock \doi{10.1186/s13014-016-0747-y}

\bibitem{Johnstone2018}
Johnstone E, Wyatt JJ, Henry AM, et~al.
\newblock Systematic review of synthetic computed tomography generation
  methodologies for use in magnetic resonance imaging–only radiation therapy.
\newblock \emph{Int J Radiat Oncol}. 2018;\hspace{0pt}100(1):199--217.
\newblock \doi{10.1016/j.ijrobp.2017.08.043}

\bibitem{Han2017}
Han X.
\newblock {MR}-based synthetic {CT} generation using a deep convolutional
  neural network method.
\newblock \emph{Med Phys}. 2017;\hspace{0pt}44(4):1408--1419.
\newblock \doi{10.1002/mp.12155}

\bibitem{Arabi2018}
Arabi H, Dowling JA, Burgos N, et~al.
\newblock {Comparative study of algorithms for synthetic CT generation from
  MRI: Consequences for MRI-guided radiation planning in the pelvic region}.
\newblock \emph{Med Phys}. 2018;\hspace{0pt}45:5218--5233.
\newblock \doi{10.1002/mp.13187}

\bibitem{Chen2018a}
Chen S, Qin A, Zhou D, Yan D.
\newblock {Technical Note: U-net-generated synthetic CT images for magnetic
  resonance imaging-only prostate intensity-modulated radiation therapy
  treatment planning}.
\newblock \emph{Med Phys}. 2018;\hspace{0pt}45:5659--5665.
\newblock \doi{10.1002/mp.13247}

\bibitem{Maspero2018a}
Maspero M, Savenije MHF, Dinkla AM, et~al.
\newblock Dose evaluation of fast synthetic-{CT} generation using a generative
  adversarial network for general pelvis {MR}-only radiotherapy.
\newblock \emph{Phys Med Biol}. 2018;\hspace{0pt}63:1--11.
\newblock \doi{10.1088/1361-6560/aada6d}

\bibitem{Olberg2019}
Olberg S, Zhang H, Kennedy WR, et~al.
\newblock Synthetic {CT} reconstruction using a deep spatial pyramid
  convolutional framework for {MR}‐only breast radiotherapy.
\newblock \emph{Med Phys}. 2019;\hspace{0pt}\doi{10.1002/mp.13716}

\bibitem{Mostafaei2018}
Mostafaei F, Tai A, Omari E, et~al.
\newblock Variations of {MRI}-assessed peristaltic motions during radiation
  therapy.
\newblock \emph{PLoS One}. 2018;\hspace{0pt}13:e0205917.
\newblock \doi{10.1371/journal.pone.0205917}

\bibitem{Nakamoto2004}
Nakamoto Y, Chin BB, Cohade C, Osman M, Tatsumi M, Wahl RL.
\newblock {PET/CT}: artifacts caused by bowel motion.
\newblock \emph{Nucl Med Commun}. 2004;\hspace{0pt}25:221--225

\bibitem{Feng2009}
Feng M, Balter J, Normolle D, et~al.
\newblock Characterization of pancreatic tumor motion using cine {MRI}:
  surrogates for tumor position should be used with caution.
\newblock \emph{Int J Radiat Oncol Biol Phys}. 2009;\hspace{0pt}74:884--891.
\newblock \doi{10.1016/j.ijrobp.2009.02.003}

\bibitem{Kumagi2009}
Kumagai M, Hara R, Mori S, et~al.
\newblock Impact of intrafractional bowel gas movement on carbon ion beam dose
  distribution in pancreatic radiotherapy.
\newblock \emph{Int J Radiat Oncol Biol Phys}. 2009;\hspace{0pt}73:1276--1281.
\newblock \doi{10.1016/j.ijrobp.2008.10.055}

\bibitem{Corradini2019}
Corradini S, Alongi F, Andratschke N, et~al.
\newblock {MR}-guidance in clinical reality: current treatment challenges and
  future perspectives.
\newblock \emph{Radiat Oncol}. 2019;\hspace{0pt}14:92.
\newblock \doi{10.1186/s13014-019-1308-y}

\bibitem{Henke2018}
Henke L, Kashani R, Robinson C, et~al.
\newblock Phase {I} trial of stereotactic {MR}-guided online adaptive radiation
  therapy ({SMART}) for the treatment of oligometastatic or unresectable
  primary malignancies of the abdomen.
\newblock \emph{Radiother Oncol}. 2018;\hspace{0pt}126(3):519--526.
\newblock \doi{10.1016/j.radonc.2017.11.032}

\bibitem{Zhu2018}
Zhu JY, Park T, Isola P, Efros AA.
\newblock Unpaired image-to-image translation using cycle-consistent
  adversarial networks.
\newblock \emph{ArXiv}. 2018;\hspace{0pt}:1--18

\bibitem{Peng2020}
Peng Y, Chen S, Qin A, et~al.
\newblock Magnetic resonance-based synthetic computed tomography images
  generated using generative adversarial networks for nasopharyngeal carcinoma
  radiotherapy treatment planning.
\newblock \emph{Radiother Oncol}. 2020;\hspace{0pt}150:217--224.
\newblock \doi{10.1016/j.radonc.2020.06.049}

\bibitem{Fu2020}
Fu J, Singhrao K, Cao M, et~al.
\newblock Generation of abdominal synthetic {CT}s from {0.35T MR} images using
  generative adversarial networks for {MR}-only liver radiotherapy.
\newblock \emph{Biomed Phys Eng Express}. 2020;\hspace{0pt}6:015033.
\newblock \doi{10.1088/2057-1976/ab6e1f}

\bibitem{Bredfeldt2017}
Bredfeldt JS, Liu L, Feng M, Cao Y, Balter JM.
\newblock Synthetic {CT} for {MRI}-based liver stereotactic body radiotherapy
  treatment planning.
\newblock \emph{Phys Med Biol}. 2017;\hspace{0pt}62:2922--2934.
\newblock \doi{10.1088/1361-6560/aa5059}

\bibitem{Hsu2019}
Hsu SH, Peng Q, Tom{\'{e}} WA.
\newblock On the generation of synthetic {CT} for a {MRI}-only radiation
  therapy workflow for the abdomen.
\newblock \emph{J Phys Conf Ser}. 2019;\hspace{0pt}1154:1--4.
\newblock \doi{10.1088/1742-6596/1154/1/012011}

\bibitem{Ahunbay2019}
Ahunbay EE, Thapa R, Chen X, Paulson E, Li XA.
\newblock A technique to rapidly generate synthetic computed tomography for
  magnetic resonance imaging-guided online adaptive replanning: An exploratory
  study.
\newblock \emph{Int J Radiat Oncol}. 2019;\hspace{0pt}103(5):1261--1270.
\newblock \doi{10.1016/j.ijrobp.2018.12.008}

\bibitem{Guerreiro2019}
Guerreiro F, Koivula L, Seravalli E, et~al.
\newblock Feasibility of {MRI}-only photon and proton dose calculations for
  pediatric patients with abdominal tumors.
\newblock \emph{Phys Med Biol}. 2019;\hspace{0pt}64:1--13.
\newblock \doi{10.1088/1361-6560/ab0095}

\bibitem{Korhonen2014}
Korhonen J, Kapanen M, Keyril{\"{a}}inen J, Sepp{\"{a}}l{\"{a}} T, Tenhunen M.
\newblock A dual model {HU} conversion from {MRI} intensity values within and
  outside of bone segment for {MRI}-based radiotherapy treatment planning of
  prostate cancer.
\newblock \emph{Med Phys}. 2014;\hspace{0pt}41(1):1--12.
\newblock \doi{10.1118/1.4842575}

\bibitem{Koivula2016}
Koivula L, Wee L, Korhonen J.
\newblock Feasibility of {MRI}-only treatment planning for proton therapy in
  brain and prostate cancers: Dose calculation accuracy in substitute {CT}
  images.
\newblock \emph{Med Phys}. 2016;\hspace{0pt}43(8):4643--4642.
\newblock \doi{10.1118/1.4958677}

\bibitem{Koivula2017}
Koivula L, Kapanen M, Sepp{\"{a}}l{\"{a}} T, et~al.
\newblock Intensity-based dual model method for generation of synthetic {CT}
  images from standard {T}2-weighted {MR} images – generalized technique for
  four different {MR} scanners.
\newblock \emph{Radiother Oncol}. 2017;\hspace{0pt}125(3):411--419.
\newblock \doi{10.1016/j.radonc.2017.10.011}

\bibitem{tf2015}
Abadi M, Agarwal A, Barham P, et~al.
\newblock {TensorFlow}: Large-scale machine learning on heterogeneous systems.
  2015.
\newblock Software available from tensorflow.org

\bibitem{Otsu1979}
Otsu N.
\newblock A threshold selection method from gray-level histograms.
\newblock \emph{IEEE Trans Syst Man Cybern}. 1979;\hspace{0pt}9:62--66

\bibitem{Vincent1993}
Vincent L.
\newblock Morphological grayscale reconstruction in image analysis:
  Applications and efficient algorithms.
\newblock \emph{IEEE Trans Image Process}. 1993;\hspace{0pt}2(2):176--201.
\newblock \doi{10.1109/83.217222}

\bibitem{Schonlieb2015}
Sch{\"o}nlieb CB.
\newblock \emph{Partial Differential Equation Methods for Image Inpainting}.
\newblock Cambridge, {UK}: {Cambridge University Press}. 2015.
\newblock \doi{10.1017/CBO9780511734304}

\bibitem{Parisotto2016}
Parisotto S, Sch{\"o}nlieb C.
\newblock {MATLAB}/{Python} codes for the {Image} {Inpainting} {Problem}.
\newblock GitHub repository, {MATLAB} Central File Exchange. 2016

\bibitem{Shi2013}
Shi J.
\newblock \emph{Statistical dependence measure based multi-modal image
  registration and registration assisted non-parametric image segmentation}.
\newblock Ph.D. thesis. University of {Florida}. 2013

\bibitem{Kingma2017}
Kingma D, Ba J.
\newblock Adam: A method for stochastic optimization.
\newblock \emph{ArXiv}. 2017;\hspace{0pt}:1--15

\bibitem{Jegou2017}
J{\'e}gou S, Drozdzal M, Vazquez D, Romero A, Bengio Y.
\newblock The one hundred layers tiramisu: Fully convolutional {DenseNets} for
  semantic segmentation.
\newblock \emph{ArXiv}. 2017;\hspace{0pt}:1--9

\bibitem{Ronneberger2015}
Ronneberger O, Fischer P, Brox T.
\newblock U-net: Convolutional networks for biomedical image segmentation.
\newblock \emph{ArXiv}. 2015;\hspace{0pt}:1--8

\bibitem{He2015}
He K, Zhang X, Ren S, Sun J.
\newblock Deep residual learning for image recognition.
\newblock \emph{ArXiv}. 2015;\hspace{0pt}:1--12

\bibitem{Low1998}
Low DA, Harms WB, Mutic S, Purdy JA.
\newblock A technique for the quantitative evaluation of dose distributions.
\newblock \emph{Med Phys}. 1998;\hspace{0pt}25(5):656--661.
\newblock \doi{10.1118/1.598248}

\bibitem{Wilcoxon1945}
Wilcoxon F.
\newblock Individual comparisons by ranking methods.
\newblock \emph{Biometrics Bulletin}. 1945;\hspace{0pt}1:80--83

\bibitem{Fu2018}
Fu Y, Mazur TR, Wu X, et~al.
\newblock A novel {MRI} segmentation method using {CNN}-based correction
  network for {MRI}-guided adaptive radiotherapy.
\newblock \emph{Med Phys}. 2018;\hspace{0pt}\doi{10.1002/mp.13221}

\bibitem{Henke2018b}
Henke L, Contreras J, Green O, et~al.
\newblock Magnetic resonance image-guided radiotherapy ({MRIgRT}): a 4.5-{Y}ear
  clinical experience.
\newblock \emph{Clin Oncol (R Coll Radiol)}. 2018;\hspace{0pt}30(11):720--727.
\newblock \doi{10.1016/j.clon.2018.08.010}

\bibitem{Ginn2017}
Ginn J, Agazaryan N, Cao M, et~al.
\newblock Characterization of spatial distortion in a 0.35 {T MRI}-guided
  radiotherapy system.
\newblock \emph{Phys Med Biol}. 2017;\hspace{0pt}62:4525--4540.
\newblock \doi{10.1088/1361-6560/aa6e1a}

\bibitem{Green2018}
Green O, Henke L, Parikh P, Roach M, Michalski J, Gach H.
\newblock Practical implications of ferromagnetic artifacts in low-field
  {MRI}-guided radiotherapy.
\newblock \emph{Cureus}. 2018;\hspace{0pt}10:e2359.
\newblock \doi{10.7759/cureus.2359}

\end{thebibliography}

\end{singlespace}

\end{document}